# An Efficient *Probe*-Based Routing for Content-Centric Networking

**Pei-Hsuan Tsai [1],\*, Junbin Zhang [2] and Meng-Hsun Tsai [2,]\***

[1] Institute of Manufacturing Information and Systems, National Cheng Kung University, Tainan City 701, Taiwan

[2] Department of Computer Science and Information Engineering, National Cheng Kung University, Tainan City 701, Taiwan; p78083025@ncku.edu.tw

\* Correspondence: phtsai@mail.ncku.edu.tw (P.-H.T.); tsaimh@csie.ncku.edu.tw (M.-H.T.)

**Abstract:** With the development of new technologies and applications, such as the Internet of Things, smart cities, 5G, and edge computing, traditional Internet Protocol-based (IP-based) networks have been exposed as having many problems. Information-Centric Networking (ICN), Named Data Networking (NDN), and Content-Centric Networking (CCN) are therefore proposed as an alternative for future networks. However, unlike IP-based networks, CCN routing is non-deterministic and difficult to optimize due to frequent in-network caching replacement. This paper presents a novel *probe*-based routing algorithm that explores real-time in-network caching to ensure the routing table storing the optimal paths to the nearest content provider is up to date. Effective *probe*-selections, Pending Interest Table (PIT) *probe*, and Forwarding Information Base (FIB) *probe* are discussed and analyzed by simulation with different performance measurements. Compared with the basic CCN, in terms of qualitative analysis, the additional computational overhead of our approach is $O(N_{CS} + N_{rt} + N_{FIB} * N_{SPT})$ and $O(N_{FIB})$ on processing interest packets and data packets, respectively. However, in terms of quantitative analysis, our approach reduces the number of timeout interests by 6% and the average response time by 0.6 s. Furthermore, although basic CCN and our approach belong to the same Quality of Service (QoS) category, our approach outperforms basic CCN in terms of real values. Additionally, our *probe*-based approach performs better than RECIF+PIF and EEGPR. Owing to speedup FIB updating by probes, our approach provides more reliable interest packet routing when accounting for router failures. In summary, the results demonstrate that compared to basic CCN, our *probe*-based routing approach raises FIB accuracy and reduces network congestion and response time, resulting in efficient routing.

**Keywords:** content-centric networking; content exploration; information-centric networking; named data networking; routing; next generation networking

## 1. Introduction

With the ubiquity of mobile devices and the increasing speed of network connections, there have been significant changes in Internet usage. Rather than providing fundamental connectivity, the Internet has essentially evolved into largely a distribution network with vast volumes of social and media content flowing from content providers to consumers [1]. In Internet communication, there has been a significant paradigm shift from current service-oriented to content-oriented [1–3]. Instead of knowing where the content provider is, today's Internet users are more interested in fast, efficient, and secure content retrieval [1,4,5]. As a result, architectures for Information-Centric Networking (ICN) have been proposed and widely developed [2,3]. Content-Centric Networking (CCN) is the most potential and innovative of all the ICN-based networks [6,7]. CCN is a name-based protocol that employs the content name as the search target rather than the IP address as the network routing destination. It can be applied to different scenarios, such as high-speed data centers, the Internet of Things (IoT), mobile edge computing, and sensor networks with limited resources. Compared to traditional Transmission Control Protocol/Internet Protocol (TCP/IP) communication protocol, CCN is more flexible and scalable by allowing data to be cached in routers and using interest with content name, which is the name of the required data, as the query packet [8]. The response time of interest can be significantly reduced by utilizing efficient routing and in-network caching [9].

The traditional TCP/IP routing protocol is a single-source to a single-destination transmission that only considers one server as the content provider and uses the shortest length or lowest cost as the path planning goal function. In this paper, the content provider is defined as the server or router that owns the content that the user requires. However, with the increasing number of IoT applications, TCP/IP networks face an insufficient number of IP addresses or require a longer IP address to handle numerous amounts of devices, putting a strain on the router. Furthermore, TCP/IP networks cannot satisfy more and more Quality of Service (QoS) criteria, such as real-time traffic (such as video confer-encing) and best-effort traffic (such as file transfer protocol (FTP)) [10]. Different from the TCP/IP protocol that uses IP to identify a single content provider, CCN uses the properties of in-network caching that scatters contents over routers to consider all routers in the network as potential content providers. CCN relies on two tables for routing: Forwarding Information Base (FIB) and Pending Interest Table (PIT). The outgoing interface of packet forwarding is determined by routers using FIB. Therefore, FIB is designed to keep track of potential content providers. PIT is used by routers to track the forwarded path of an unsatisfied interest so that data can be sent back along that path, as well as to prevent duplicate forwarding of interests requesting the same content. However, Basic CCN's planning routing is inefficient [3]. Because of insufficient FIB space or out-of-date FIB entries, basic CCN experiences frequent retransmission and network congestion. As a result, many studies propose exploring the distribution of cached content to improve the CCN. Based on how to obtain cache information, they are categorized into content share [5,11–13] and content query [3,4,7]. Content share is that the router periodically shares cache information with other routers. Content query is that the router asks for other routers' cache information. However, according to our observations or surveys, their common shortcoming is extra-packets for exploring the distribution of cached content which significantly increases network traffic overhead.

To tackle this issue, this paper presents an efficient *probe*-based routing approach for CCN. So long as the router's FIB is correct and up-to-date, it is possible to choose the best content provider. Therefore, our approach focuses on the FIB updating algorithm. A path planning table is also added in our approach for determining the content provider. Instead of the outgoing interface, FIB is modified to record the candidate content providers. The shortest path to content providers is stored in the path planning table. Many TCP/IP routing algorithms can be used [14,15].

Our approach adds a new field called *probe* to the interest and data packets to allow for cache content exploration. The *probe* that is utilized to update FIB can be any content in the network. Hence, one key ingredient in the FIB updating algorithm is *probe* selection. The selection of *probe* can be considered from two aspects, depending on the different performance measures. One goal is to satisfy pending interests as much as feasible to directly reduce interest waiting time, so the *probe* is selected from PIT, called *PIT probe*. The other is to improve routing accuracy to reduce interest retransmission, so the *probe* is selected from FIB, called *FIB probe*. In this paper, *PIT probe* selects the most popular entry from PIT as the *probe*. *FIB probe* selects the entry with the highest cost or the least updated as the *probe*. Our approach is an effective optimization as compared to basic CNN, as evidenced by the findings. *PIT probe* performs best in terms of waiting time, whereas *FIB probe* performs better in all the other measurements.

In summary, the contributions of this paper include the following:
- This paper presents an efficient probe-based routing algorithm that hybrid the advantages of TCP/IP and CCN.
- Compared to basic CCN, our approach improves the routing mechanism to find the best content provider and the shortest path to the nearest content by modifying FIB, interest packet, and data packet, as well as adding a path planning table. By real-time exploring cached content in routers and finding the shortest path to the nearest content, our approach reduces network congestion, reduces response time, and raises FIB accuracy. Compared to other content exploration methods that relied on extra designated packets that increase the network burden, our approach minimizes the cost of exploration.
- Different *probe* selections, such as from PIT or FIB, are compared and analyzed. Except for the waiting time of pending interests, *PIT probe* performs better on average response time, while *FIB probe* has a better performance in all other measurements, which shows that the best content provider optimization is effective.
- Compared to RECIF+PIF and EEGPR, owing to speedup FIB updating by probes, our approach provides more reliable interest packet routing when the router failures.
- A real network topology is simulated to discuss the feasibility of our approach. The results show that it is highly compatible with the existing cache replacement strategy and forwarding strategy.

The rest of this paper includes the following sections: Section 2 introduces the traditional IP-based routing protocols, the basic CCN background, and references about content exploration in CCN. Section 3 introduces the routing process of basic CCN and then describes our approach, including data structure, *probe*-based routing algorithms, and routing complexity. Section 4 introduces the simulation platform setup and the results are discussed in detail. Our work is summarized in Section 5.

## 2. Related Work

### 2.1. Traditional IP-Based Routing

In existing networks, IP-based routing protocols are commonly employed. Many algorithms, such as shortest path routing [14], have been developed to optimize routing to reduce network delay. However, the limited IP address is insufficient for an increasing number of IoT device applications. Many studies and protocols have been proposed to solve the problem [16–18]. By altering the network address information in the IP header of the data packet, network address translation (NAT) maps one IP address space to another [17]. Classless Inter-Domain Routing (CIDR) is a new method of assigning IP addresses that have replaced the old Internet classified network addressing architecture [18]. To overcome this issue, IPv6 increases the IP address size from 32 bits to 128 bits [16]. They still do not cover all IP-based network applications, such as less powerful IoT and sensor networks.

*2.2. Basic CCN*

In-network caching is one significant feature of CCN, where content is scattered and cached among routers. Different from an IP-based network, CCN uses name-based routing that retrieves contents from nearby router caches. Multipath forwarding is also supported by CCN to reduce response time. Between the application layer and transportation layer, the CCN protocol alters IP-based protocol by adding three layers: a security layer, a named contents layer, and a strategy layer [19].

*Interest packets* and *data packets* are two types of packets in basic CCN [20–23]. When receiving an interest packet, the router first tries to retrieve *Name* from its Content Store (CS) and return the data packet. In the router, CS is used to cache network content. If the CS retrieval fails, the router adds the interest packet to the PIT. The information stored in PIT includes *Name*, *incoming interface*, and *outgoing interface*. The following interest request for a duplicate *Name* is aggregated to the same entry as if it was in PIT [7]. FIB contains entries with *Name* and *outgoing interface*. There are multiple outgoing interfaces for each unique *Name*. Many forwarding strategies [3–5,7,24–26] have been proposed, such as broadcast, which forwards interest through all outgoing interfaces, and best route, which forwards interest through one outgoing interface at a time starting with the lowest cost.

*2.3. Extensions of CCN*

Recent CCN-related studies have focused on improving forwarding efficiency [25,27–29]. Rosensweig and Kurose [27] describes an implicit coordination approach in which a Breadcrumb is used to record the direction and time of the request forwarded in the past. SCAN, as proposed in [28], is a scalable content routing. During the forwarding process, it sequentially queries the surrounding neighbors to obtain a copy of the requested content. Hoque et al. [25] designed a routing extension to Open Shortest Path First (OSPF) for NDN, called OSPFN. They defined a link-state announcement (LSA) to carry the content name to build FIB. OSPFN is confined to a single path because it still uses the IP address as the router ID. Therefore, they developed a Named-data Link State Routing protocol (NLSR) to solve the OSPFN problem [25]. Eum et al. [29] introduced the Potential Based Routing (PBR) to tackle namely routing problem. Each router actively broadcasts its cached content to maintain a table of potentially reachable copy information. Siddiqa et al. [30] proposed a novel Energy Efficient Greedy Routing Protocol (EEGRP). It minimizes flooding by using uniquely designed routing table (RT) entries in the network. The provider node selects the shortest path to forward data packets within the minimum delay to achieve high energy efficiency. Yang et al. [31] proposed an Intelligent CCN routing strategy based on Bacterial Quorum pattern (ICBQ). It simulates the quorum sensing behavior of bacteria to obtain parameter information such as bandwidth, delay, error rate, and adaptive chemotaxis behavior to select the optimal interface to forward packets. Cai et al. [32] presented a Software Defined Status Aware routing (SD-STAR) method that combined CCN and SDN. The controller is used by this mechanism to collect network statuses, then update the interest packet, choose the best forwarding interface, and choose the best routing path.

*2.4. Content Exploration in CCN*

One solution to fully exploit in-network cache is to explore the distribution of cached content [3–5,7,24,25,33]. They can be categorized into *content share* and *content query*.

*(1) Content Share* indicates that the router regularly broadcasts cache information to other routers.

Yaogong Wang et al. [5] discussed the necessity and feasibility of routers advertising cached information in the network. Jason Min Wang et al. [12] proposed an intra-Autonomous Systems (intra-AS) cache cooperation scheme. The router periodically advertises the cache "summary" of its currently cached information to neighboring routers. Zhe Li et al. [21] proposed a cooperative caching strategy to handle the distribution of large video streams. This method can make each Content Routers (CR) cache different data and avoid caching the same data on neighboring CRs. Wong et al. [11] proposed a cooperative scheme that is a cached-bit strategy informing other routers along the path whether a given content has already been cached in the network or not through a bit set. Saha et al. [34] and Sen Wang et al. [35] proposed a hash-based cooperative caching strategy, which uses hash values to map content names to router IDs. Sato et al. [24] also proposed an effective hash-based cache distribution and search scheme. The CS of all routers in the network is managed by an independent autonomous system. When cached content is updated, all neighboring routers are notified to update FIB by broadcast. Marandi et al. [36] proposed a Pull-based Bloom Filter-based Routing (BFR) method, which only advertises the required file name. It is better than the original BFR in terms of overhead, latency, and memory. The notion of SDN has been integrated into CCN in some studies [30,37]. To solve the problem of estimating the location of the cache server in the centralized cache management of SDN-based ICN, Badshah et al. [37] used singular value decomposition (SVD) and QR-factorization with column pivoting methods. The *probe*-based approach is designed to improve the precision of FIB, especially for the peer-to-peer distributed networks. Compared to peer-to-peer distributes networks, SDN has a controller to manage the routing tables and data storage so it is less significant to apply a probe-based approach in SDN.

*(2) Content Query* indicates that the router asks for other routers' cache information by sending an extra packet.

Raffaele et al. [4] proposed a hybrid forwarding solution to exploit the known routing information and explore the unknown cache content. They also [26] proposed a dynamic INterest FORwarding Mechanism (INFORM) based on a reinforcement learning framework that is a distributed Q-leaning. Rossini et al. [38] proposed an ideal Nearest Replica Routing (iNRR) forwarding policy. This policy can explore a meta-request that only needs to respond to true or false. Bastos et al. [3] proposed a Diversity-based Search-and-Routing (DIVER) approach, which uses a probe packet with ε probability to search for copies of cached content on surrounding routers. To make full use of multipath forwarding and in-network caching, Xiaoyan et al. [7,33] proposed an on-demand off-path Cache Exploration-based Multipath Forwarding strategy (O2CEMF).

However, extra-packet is the significant burden for all of the above methods, which increases redundant traffic, computational complexity, and network overhead [1,3,7]. Moreover, several content query approaches have an additional table to manage the candidate exploration object. Compared to those approaches, our approach fully utilizes the advantages of in-network cache and effectively improves these problems.

**3. Design and Description of Our Proposal**

This section illustrates how our approach improves the basic CCN in terms of both data structure and routing algorithm.

*3.1. Data Structure*

3.1.1. Routing Tables

- *FIB*: In our design, FIB is modified to keep track of the potential content providers. In the FIB of a basic CCN, the outgoing interface is replaced to hold potential content providers by router ID. Figure 1 demonstrates our FIB table.

| Content Name | Router Name | |
|---|---|---|
| name | router ID | ... |
| ... | ... | ... |

**Figure 1.** FIB structure.

- *PIT*: It has the same definition as basic CCN.
- *Shortest Path Table* was designed to optimize routing to decrease network delay. As shown in Figure 2, the fields of Shortest Path Table (SPT) include *Destination*, *Cost*, and *Outgoing Interface*. *Destination* is the router ID. *Cost* obtaining content from the designated router might be measured by response time, hop count, or other metrics. *Outgoing Interface* is the outgoing interface to a certain destination. In this paper, the path is calculated using a dynamic routing algorithm (OSPF), which is described in Section 3.2.

| Destination | Cost | Outgoing Interface |
|---|---|---|
| router ID | cost | outgoing interface |
| ... | ... | ... |

**Figure 2.** Shortest path table structure.

3.1.2. UP-To Date FIB Mechanism

Compared to other content explorations that periodically and completely send designated packets to explore the distribution of cached content, our approach improves the diversity and miss rate of FIB with minimal network overhead by appending fields to conventional interests and data packets. Two fields, *probe* and *probe-response*, are added to both the interest packet and the data packet, as shown in Figure 3a,b.

- *Probe* is a content name. Depending on the performance measurement, the *probe* can be arbitrarily assigned. In this paper, according to the QoS, the *probe* is chosen from either PIT or FIB. *PIT probe* selects the most popular entry in PIT to satisfy as many pending interests as feasible and further directly reduce interest waiting time. *FIB probe* selects the entry with maximum cost from FIB to increase routing accuracy and prevent transmission failure. Section 4.4 simulates and analyzes the two selections on different performance measurements.
- *Probe-response* is a list of the *probe's* content providers. The router identification (ID) is added to the list when the *probe* is matched during the exploration phase. Routers update their FIB using *probe-response* data.

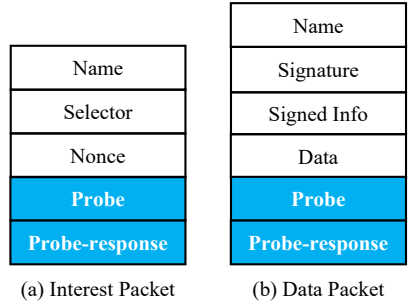

(a) Interest Packet  (b) Data Packet

**Figure 3.** Packet format.

*3.2. Probe-Based Routing Algorithms*

The routing procedure is described as follows. A router's routing is triggered by packets. The *interest sending phase* and the *data packet returning phase* are the two transmission phases.

3.2.1. During interest sending phase

In basic CCN, when a router receives an interest, it performs three steps:

**Step 1**: Check its CS. If the *Name* matches in CS, the content data is returned as a data packet via the interest's incoming interface. Otherwise, go to step 2.

**Step 2**: Check its PIT. If the *Name* matches in PIT, the incoming interface is aggregated into the same entry and the interest is dropped. Otherwise, PIT creates a new entry with the *Name* of the interest and then goes to step 3.

**Step 3**: Check its FIB. If the *Name* matches in FIB, the interest is forwarded through the minimum cost of the matched outgoing interface. Otherwise, the interest is broadcast by the routers.

The adjustments of basic CCN in this paper are described as follows. Figure 4 illustrates an example of our *probe*-based approach. When a router receives an interest, it can play one of three roles: Initial (I), Miss (M), or Hit (H). If the interest is sent from the consumer, the router is called *I router*. If the router fails to retrieve data of *Name* from its CS, it is called *M router*. If the router owns the data of *Name*, it is called *H router*.

*I router* adds a *probe* to interest before forwarding it. The *probe* is selected from one of two options: PIT or FIB (see lines ① and ② in Figure 4). *I router* then seeks the best candidate content provider with the minimum cost in SPT, stored in the matched FIB entry in Figure 1 (see line ③ in Figure 4). To forward the interest, the outgoing interface to the best candidate content provider is selected (see line ④ in Figure 4).

*M router*, in addition to the basic CCN routing steps, checks if it is the *probe's* content provider. If affirmative, the router ID is added to *probe-response* (see line ⑤ in Figure 4) and then the interest is forwarded as *I router* without adding a new *probe* (see line ③ and ④ in Figure 4).

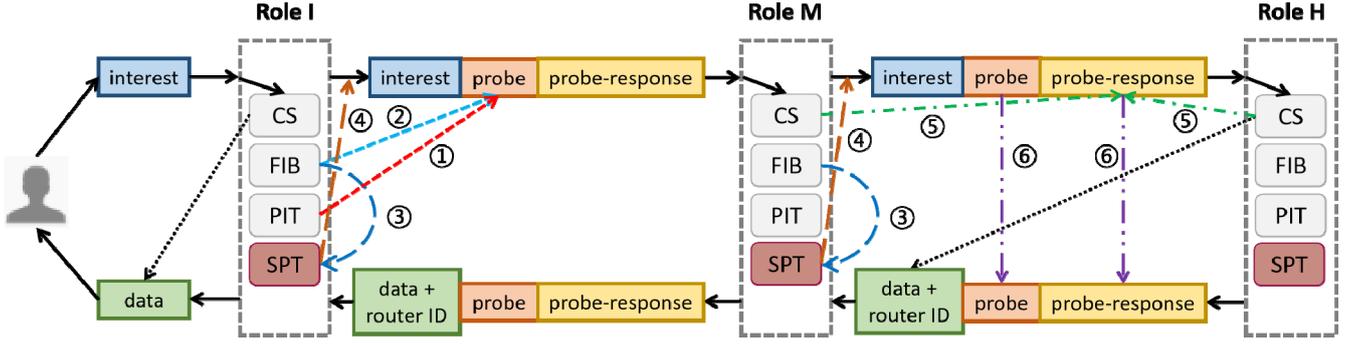

**Figure 4.** *Probe*-based routing process.

*H router* adds its ID to the data packet. In addition to the *M router* steps, the *probe* and *probe-response* in the interest are replicated to the data packet (see line ⑥ in Figure 4).

3.2.2. During data packet returning phase

When a router receives a data packet in basic CCN, it first checks to see if there is a matching PIT entry. If the *Name* matches, the router forwards the data packet to all of the routers listed in the PIT entry before deleting it. Otherwise, because it implies that the same data packet has been received, the data packet is dropped. Meanwhile, the router may cache the data depending on its cache replacement policy and update FIB.

At this phase, our approach improves basic CCN by updating FIB as described below.

**FIB**: In the data packet, there are two *Names*: *data name* and *probe name*. The *probe name*, in addition to the *data name*, is used to update FIB. If the *Names* were already stored in FIB, the matched entry appends the router IDs to the data packets. Otherwise, a new FIB entry is created accordingly. Since the FIB entry's space is limited, a replacement policy is necessary. Generally, a time-based replacement policy, such as Least Recently Used (LRU), Least Frequently Used (LFU), First In First Out (FIFO), or a cost-based replacement policy, such as shortest path, QoS, is adopted.

**SPT**: There are two types of networks: static and dynamic, based on network traffic. The transmission time between any two connective routers is identical in a static network, while the transmission time between any two connective routers is different in a dynamic network. In a static network, the hop count might be considered a cost. The SPT is sustainable as long as the network topology is static and the shortest paths between any two routers are fixed. In a dynamic network, the cost can be traveling time or other metrics that are impacted by network conditions (such as congestion). Many traditional network routing algorithms, such as Open Shortest Path First (OSPF) [14], can be used to optimize SPT. OSPF is a TCP/IP network routing protocol that is dynamic, hierarchical, and Link–State [39]. OSPF uses Dijkstra's shortest path algorithm to find the best router [15]. In this paper, the network is static and the cost is the number of hops. The SPT is only updated when the network topology changes.

*3.3. Routing Complexity*

For network overhead, it includes bandwidth overhead and time overhead.

3.3.1. Bandwidth Overhead

In our probe-based routing approach, for each interest packet and data packet, the additional overhead is 2 bytes to store a content name and 20 bytes to store a router ID. As described in Section 4.2, as compared to the basic CCN, our approach adds the size of 22 bytes for interest packets and data packets. As a result, our approach's bandwidth overhead is k ∗ 22 bytes, where k is the throughput.

3.3.2. Time Overhead

Algorithms 1 and 2 demonstrate our approach of processing interest packets and data packets, respectively, which were modified from basic CCN. The following analyzes the additional time complexity caused by our modifications. Algorithm 1 has three additional parts compared to CCN, which are as follows:
(1) Check the probe in CS (see lines 7–9).
(2) Add a probe (see lines 13–18).
(3) Select outgoing interfaces for interest packets (see lines 19-24).

**Algorithm 1** Processing Interest Packet

    **OnInterest** (*interest*)

1  // Check PIT entries for matching *interest*. 0 means not found.
    **if** PIT (*interest*) == 0 **then**
2    Create a PIT entry for *interest*
3  **else** merge *interest* to PIT, drop *interest* and **break**
4  **end if**
5  *data* = Create Data ()
6  // Check CS entries for matching *interest*. 1 means found.
    **if** CS (*interest*) !=0 **then**
7    // Check CS entries for matching *probe*.
    **if** CS (*interest.probe*) !=0 **then**
8      add routerID to *probe-response*
9      add *probe-response* to *data*
10   add content to *data*
11  **Return** *data*
12  **end** *if*
13  // If no matching *interest* in CS, then continue to check CS entries for matching probe.
    *else if* CS (*interest.probe*) !=0 then
14      add routerID to *probe-response*
15  // Check if *interest* needs to add a *probe*.
    *else if interest.probe* == 0 **then**
16      // select a *probe* in PIT or FIB
      name = find an entry in PIT (max popular) or FIB (max cost)
17      add name to *interest.probe*
18  **end** *if*
19  // outgoing interface = FIB (min cost)
20  // Get routerID sequentially from FIB entry
    for each routerID in FIB:
21    // Get the cost of routerID from SPT and find routerID with lowest cost
    *if* SPT (routerID) < mincost
22      mincost = SPT (routerID)
23      outgoing interface = routerID
24    **end** *if*
25  **Forward** *interest*
26  **end** *if*
27  **end** *if*

The worst-case time complexity of checking the probe in CS is O($N_{CS}$), where $N_{CS}$ is the cache size of CS. The time complexity of the probe selection policy is O($N_{rt}$), where $N_{rt}$ is the size of the routing table. In this paper, $N_{rt}$ denotes either the FIB or PIT size, as indicated in line 16. Line 19 is the basic CCN outgoing interface selection, with a time complexity of O($N_{FIB}$), where $N_{FIB}$ is the FIB cache size. Line 20–24 is the modified outgoing interface selection of our approach. The time complexity is O($N_{FIB} * N_{SPT}$). Our approach traverses the SPT in O($N_{SPT}$), where $N_{SPT}$ is the number of entries. SPT is to record the cost and interface with the shortest path to each other router in the network so that the number of entries equals the number of routers $N_R$.

**Algorithm 2** Processing Data Packet

    **OnData** (*data*)

1  // Check whether the *probe-response* is empty.

```
       if data.probe-response !=0 then
          // Check the probe in FIB
   2      if FIB (data.probe) ==0 then
   3         Create a FIB entry for data.probe
   4      else Update the entries in FIB by data.probe-response
   5      end if
   6   end if
   7   if PIT (data) !=0 then
   8      if CS (data) ==0 then
   9         Cache data
  10      end if
  11      if FIB (data) ==0 then
  12         Create a FIB entry for data
  13      else Update the entries in FIB by data
  14      end if
  15      Forward data, remove the entry named data in PIT and break
  16   else drop data
  17   end if
```

In comparison to CCN, Algorithm 2 has one additional component: updating FIB via *probe-response* (see lines 1-6). The worst-case time complexity for checking the probe in FIB is O($N_{FIB}$). For qualitative analysis, compared to the basic CCN, the additional time complexity of our approach is O($N_{CS} + N_{rt} + N_{FIB} * N_{SPT}$) and O($N_{FIB}$) on processing interest packets and data packets, respectively. In addition to online time complexity, there is an offline Breadth-First Search (BFS) that is used to establish SPT and has a time complexity of O($N_R + N_E$), where $N_E$ is the number of network edges. However, our experiments demonstrate that our approach outperforms the basic CCN in terms of the number of timeout interest packets and average response time. The details are described in Section 4.4.4.

## 4. Performance Evaluation

### 4.1. Platform Setup

The simulator designed in this paper was implemented in Python language. The simulator was built on a computer with an Intel Xeon E5-1620 3.6 GHz processor, 32 GB of memory, and the Ubuntu 18.04 operating system. All the simulations in this paper were implemented under this platform.

### 4.2. Simulation Parameters

In the simulations, an open network topology with 12 city nodes, Abilene topology [40] is adopted as illustrated in Figure 5a. Each node is a router with both a producer and a consumer configuration. Each producer (router) generates 100 unique contents with its name prefix and numbering from 0 to 99. Taking the router "Atlanta" as an example, the contents are "Atlanta/0", ..., "Atlanta/99". The Abilene network topology contains 12 routers so the total number of contents is 1200. Each consumer randomly selects a content name from the 1200 contents to send the interest packet. In other words, the contents are chosen with the same uniform probability. Consumers send new interest packets to the network at a frequency of 1 per second.

In addition, a more complicated network topology, PoP-level Rocketfuel SPRINT topology [7,33], as shown in Figure 5b is also adopted for simulation. The network topology has 52 nodes, including 8 producers (red stars), 11 consumers (green squares), and 33 routers (blue circle). Each producer generates 200 unique contents with its name prefix, such as "R9/0", ..., "R9/199", and so on. Therefore, the total number of contents is 1600.

For basic CCN, the interest packet and data packet sizes are 5 bytes and 133 bytes respectively. The *probe* field is a 2 byte string. The *probe-response* field is recording the router IDs. The router ID is a 4 byte int. Therefore, the *probe-response* field is $n * 4$ bytes in size. In this paper, the *probe-response* only records the nearest 5 potential content providers, so n is 5. In Abilene topology, to see the impact of cache size and cache update ratio, increasing cache size ratio ranging from 1% to 40% and cache update ratio ranging from 1% to 50% are used. In PoP-level SPRINT topology, the bandwidth of the emulator is set to 1024 bits. To evaluate the impact of traffic variations on the performance of our approach, we changed the frequency of consumers sending new interest packets. The frequency was set to send from 1 to 30 interests per second. Cumulative number of router failures was used to evaluate the impact of router failure be-havior on CCN performance. The faulty router was randomly selected from 33 routers. The replacement policy for cache and FIB was FIFO and LRU. Forwarding strategies included broadcast and best route. All the simulations ran 480 s each time. Basic CCN, EEGRP [31] and ReCIF + PIF [41] as comparison methods adopted the same settings in this paper. Table 1 lists the detailed experimental parameter settings.

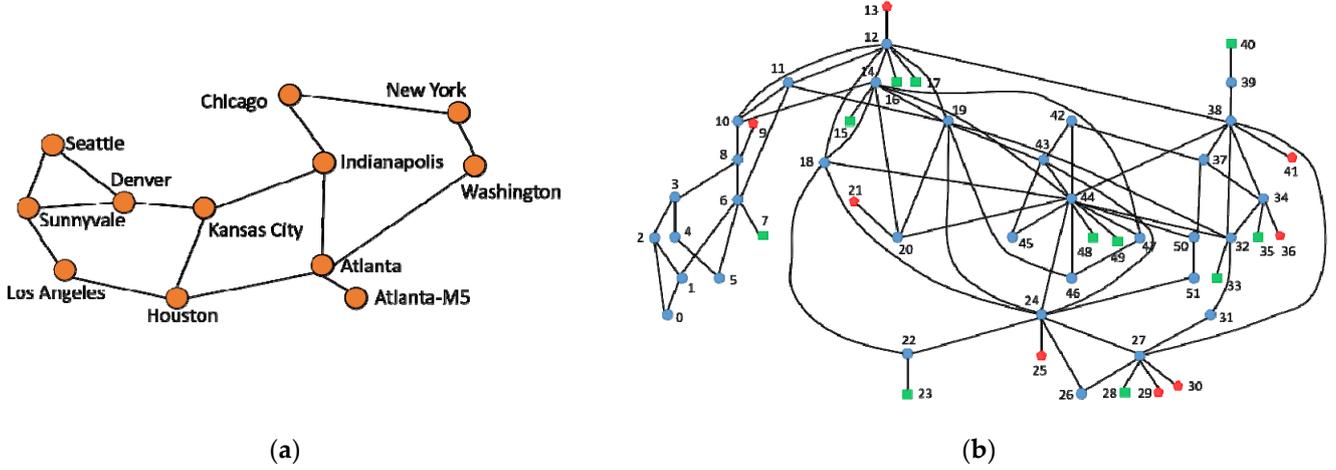

(**a**)　　　　　　　　　　　　　　　　　　　(**b**)

**Figure 5.** Topology. (**a**) Abilene topology [40]; (**b**) PoP-level SPRINT topology [7,33], where including producer (red star), consumer (green square), and router (blue circle).

**Table 1.** Experimental parameters setting.

| Parameter | Description |
| --- | --- |
| Number of routers | Abilene: 12 <br> PoP-level SPRINT: 52 |
| Propagation delay | 1 s |
| Interest packet size | Name: 2 bytes; Selector: 2 bytes; Nonce: 1 byte; Probe: 2 bytes; Probe-response: 20 bytes. |
| Data packet size | Name: 2 bytes; Signature: 2 bytes; Signed info: 1 byte; Data: 1 kb; Probe: 2 bytes; Probe-response: 20 bytes. |
| Frequency | Abilene: 1 interest/s <br> PoP-level SPRINT: 1–30 interest/s |
| Number of contents | Abilene: 1200 <br> PoP-level SPRINT: 1600 |
| Cache size ratio | 1%, 5%, 10%, 15%, 20%, 25%, 30%, 35% and 40% |
| Cache update ratio | 1%, 5%, 10%, 15%, 20%, 30%, 40% and 50% |
| Cumulative number of router failures | 1–20 |
| Cache replacement policy | FIFO and LRU |
| Forwarding strategy | broadcast and best route |
| Timeout timer of interest trails | 0.5 s |
| Simulation time | 480 s |

*4.3. Metrics*

In this paper, eight performance measurements were adopted for evaluation. They are defined as follows:

- *Effectivity of probe* is represented by the accuracy of the content provider and the routing cost. The accuracy of the content provider is defined as the ratio of data packets received by *I router* that is consistent with its expected content provider. The routing cost is defined as the number of hops a data packet passed by to reach *I router*. This metric helps to select a more efficient *probe*.
- *Total number of forwarded interests* is the amount of interests forwarded by routers. In other words, consumer-issued initial interests are not counted. This metric presents the network congestion.

- *Number of timeout interest* is the amount of PIT entries missed deadline. Each PIT entry has a deadline. The router selects another content provider in FIB as the best content provider for the time-out entry in PIT and then re-sends the interest packet. This metric is used to demonstrate the accuracy of routing table.
- *Average response time (Latency)* is the average time that takes for *I router* to get a data packet response. This metric is used to compare the routing efficiency.
- *Throughput* is the rate at which network data is successfully transmitted in a certain amount of time, and its units include bits, bytes, and packets per second [15]. In this paper, throughput is defined as the sum of all routers' received packets divided by the simulation duration.
- *Packet Loss* indicates that the packet is dropped before reaching its designated destination [15]. It is measured as a rate of packets loss compared to packets sent. This metric allows observing network collisions and congestion. The formulation is

$$Packet\ loss = \frac{(Sent\ packets - Received\ packets)}{Sent\ packets} \times 100\% \tag{1}$$

- *Delay* is the total nodal time for a packet to be transmitted from one communication endpoint to another over the network, including node processing delay, queuing delay, transmission delay, and propagation delay [42]. The more the network traffic, the longer the delay. The equation is

$$Average\ Delay = \frac{Sum\ of\ packets\ Delay}{Total\ number\ of\ packets} \tag{2}$$

- *Jitter* is used to describe the degree of network delay variation [15]. The formulation is

$$Jitter = \sqrt{\frac{\sum(Delay - Average\ Delay)^2}{Received\ packets - 1}} \tag{3}$$

*4.4. Experimental Results*

In this paper, our approach is compared with basic CCN, EEGRP [31], and ReCIF+PIF [41].

4.4.1. The Impact of Probe Selections

In this simulation, without losing generality, two additional *probe* selections are used. They are *sequential probe* that selects an entry from FIB sequentially as a *probe* and *random probe* that selects a content name from either PIT or FIB randomly as a *probe*.

Table 2 demonstrates the accuracy of the content provider and the length of the routing path, respectively.

**Table 2.** Effectivity of probe.

| Methods | Accuracy of Content Provider | Routing Path |
|---|---|---|
| PIT probe | 19.98% | 6454 |
| FIB probe | 21.39% | 6320 |
| Sequential probe | 18.99% | 6543 |
| Random probe | 18.77% | 6430 |

Our *probe*-based approaches rely on FIB and SPT to find the best content provider with the shortest path to reach it. However, the result reveals that the expected content provider is less than 20%. Because router routing tables are not centrally computed or maintained in CCN, routing trajectories with high probability are different from what *I router* expects.

The better the probe selection for our *probe*-based approach, the higher the expected content provider rate and the shorter the path length. Among the *probe* selections, *FIB probe* performs the best, because *FIB probe* is devoted to improving the highest-cost content name in FIB. However, since *PIT probe* can reduce the waiting time of interest, *FIB probe* and *PIT probe* are used in the following simulations.

4.4.2. The Impact of Cache Size Ratio

The cache size to whole content ratio has a huge impact on routing performance [3]. The higher the ratio, the less the rate of cache misses. Due to the increasing cache size, a router can cache more content. In other words, because there are more candidate content providers for an interest, obtaining content responses from nearby routers is easier. As a result, the number of forwarded interests decreases, and network congestion decreases indirectly.

Figure 6 exhibits the effect of the cache size ratio on different performance measurements of basic CCN and our *probe*-based approach. According to our observations, when the cache size ratio is more than 50%, the routing results between the basic CCN and our *probe*-based approach are scarcely different. The reasons above are that the *I router* or its neighbor routers satisfy the majority of interest packets, and the *probe*'s exploration in CCN is limited. Generally, the cache size ratio is rarely larger than 50% in most actual network situations. The amount of network content is exponentially increasing, especially with the recent increasing network applications. The large size of cached contents becomes a burden on a router and causes processing time to be delayed. As a result, only the results of the cache size ratio less than 45% are presented and discussed in this paper.

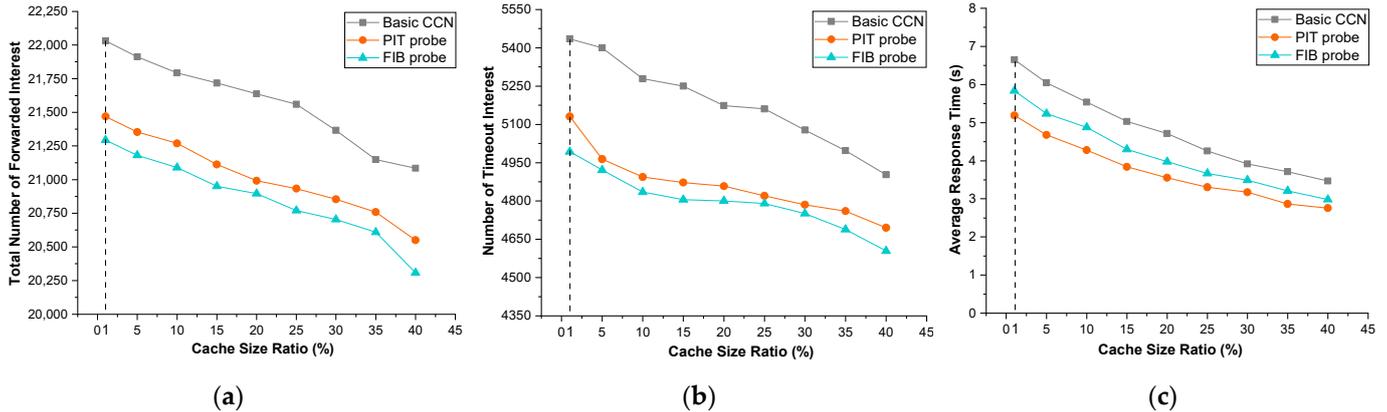

**Figure 6.** Impact of cache size ratio. (**a**) Total number of forwarded interests; (**b**) number of timeout interests; (**c**) average response time.

For the three approaches in Figure 6a–c, the number of forwarded interest packets, the number of timeout interest packets, and the average response time all decrease as the cache size ratio increases. Compared to the 1% cache size ratio, owing to increasing the cache size ratio to 40%, the number of forwarded interest packets is reduced by about 4.5% and the number of timeout interest packets by about 10%. In the simulations, the packet transmission delay in each router was set to 1 s, thus the average response time corresponds to the number of hops in the routing path. The average response time is roughly 3 to 5 s when the cache size ratio is 40%. In other words, the *I router* or its neighbors satisfy the majority of interests.

The results shown in Figure 6 illustrate that both the *PIT probe* and the *FIB probe* outperform basic CCN. *PIT probe* and *FIB probe* lower the number of forwarded interest packets of basic CCN by 1.024 and 1.034 times, respectively. The *PIT probe* and the *FIB probe* reduce the average response time of consumers by 1.08 s and 0.64 s, respectively. This also demonstrates that *probe*-based routing improves the accuracy of the routing table. It is evident that using a *probe*-based approach reduces basic CCN network congestion and improves the accuracy of the routing table, allowing for more efficient routing.

Figure 6 also demonstrates that *FIB probe* outperforms the other *probe*-based approaches in reducing network congestion. This agrees with the result of Table 2. Compared to *PIT probe*, *FIB probe* routing is more accurate, resulting in more interests being satisfied by the expected content provider and a lower routing cost. It is evident that adopting *FIB probe* as a FIB update algorithm improves finding the best content provider and the shortest path to the nearest content, resulting in the best performance in terms of the total number of forwarded interest packets and interest packet timeout.

However, *PIT probe* directly reduces customer waiting times by selecting the most popular entry in PIT as a *probe* to find the best content provider. Hence, the *PIT probe* gets superior performance on average response time than other approaches.

4.4.3. The Impact of Variation of Cached Contents

Variation of cached content is used to simulate different application scenarios and to evaluate the applicability of our *probe*-based approach. Consider IoT sensor data, video aggregator data, and social network data: their data contents vary fast, and consumers request sequential data, so cached data in a router is quickly replaced, resulting in a high cache miss rate. Intuitively, a high cache miss rate causes FIB to invalidate content providers so the FIB miss rate increases in tandem with the cache miss rate.

Figure 7 exhibits the effect of the cache update ratio on the performance of basic CCN and our *probe*-based approach. For all three approaches, the number of forwarded interest packets, the number of timeout packets, and the average response time gradually increase as the cache update ratio increases. Again, *probe*-based approaches perform better than basic CCN. The results of the comparison in Figure 7 are comparable to those in Figure 6.

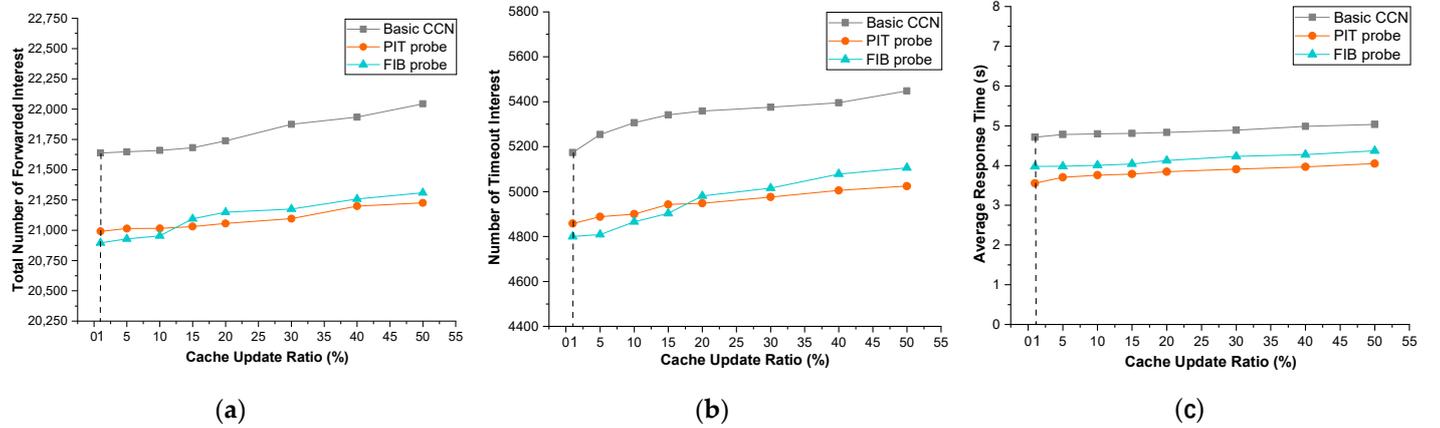

**Figure 7.** Impact of variation of cached contents. (**a**) Total number of forwarded interests; (**b**) number of timeout interests; (**c**) average response time.

The slight difference is that when the cache update ratio exceeds 15%, the *FIB probe* performs worse than the *PIT probe* in terms of the number of forwarded interest packets and timeout interest packets. This is due to the fact that the *FIB probe* updates the entry in FIB with the maximum cost yet the entry may not have been accessed recently. With the increase of time, the entry's efficacy falls dramatically, especially if the cache update ratio is high. In other words, when the FIB entry is accessed, the content providers stored in it become invalid.
On the contrary, *PIT probe* is like a greedy algorithm devoted to shortening the pending interests by selecting the most popular entry in PIT as a *probe*, which means the updated entry in FIB is accessed right away, and its effectiveness is less influenced by time. Therefore, the cache update ratio has less impact on the *PIT probe*.

4.4.4. Analysis of Routing Complexity

To quantitatively evaluate the routing complexity of our approach, the number of timeout interests and average response time are utilized as measurement parameters. In Section 3.3, the additional complexity of our approach combined with the algorithm was qualitatively explained in detail. Compared to the basic CCN, the additional time complexity of our approach are $O(N_{CS} + N_{rt} + N_{FIB} * N_{SPT})$ and $O(N_{FIB})$ on processing interest packets and data packets, respectively. However, the results in Figure 6b,c show that our approach outperforms the basic CCN in terms of the number of timeout interest packets and average response time of the consumer. *PIT probe* and *FIB probe*, respectively, reduce the number of timeout interest packets by 6.11% and 7.23%, and the average response time of consumer by 1.08 s and 0.64 s. The main reason is that the *probe*-based approach uses *probe* to find the best content providers to improve the routing table's accuracy, allowing more interests to be satisfied by nearby routers and thereby reducing interest transmission. These benefits compensate for the additional computing overhead of packet processing in each router.

4.4.5. Analysis of QoS

QoS is described as the network's use of various technologies to provide better service capabilities for specific network communications in the field of computer networking and other packet-switched telecommunication networks [43]. It is also a measurement of a network's overall performance. Especially for streaming multimedia scenes, it is very important, such as video conferences. The service capability of network communication facilities to meet end-user needs can be evaluated qualitatively and quantitatively using QoS [15,43]. For QoS analysis, the four performance metrics in Section 4.3, *throughput*, *packet loss*, *delay*, and *jitter*, are used. The QoS category is based on the literature [15]. According to the experimental results, both basic CCN and our approach are in the same QoS category. In terms of real values, however, our approach outperforms others.

Table 3 demonstrates QoS results. For throughput, both the basic CCN and our approach belong to the "Good" QoS category, because according to the literature [15], the throughput of "good" category is greater than 75 packets per second. *PIT probe* and *FIB probe* transmit two and six packets per second more than the basic CCN, respectively. This shows that our approach is better than basic CCN in improving network congestion.

Packet loss of the basic CCN and our approach also belong to the "Good" QoS category because their results are less than 15% [15]. Furthermore, packet loss of *PIT probe* and *FIB probe* is reduced by 0.86% and 1.06%, respectively, compared to the basic CCN. For throughput and packet loss, *FIB probe* performs slightly better than the basic CCN and *PIT probe*. *FIB probe* improves the accuracy of the routing table and alleviates network congestion by finding the best content provider and the shortest path to the nearest content.

**Table 3.** QoS parameters in three approaches.

| QoS Parameter | Methods | | | QoS Category |
| --- | --- | --- | --- | --- |
| | Basic CCN | PIT probe | FIB probe | |
| Throughput packets/s | 85 | 87 | 90 | Good |
| Packet loss/% | 11.4 | 10.54 | 10.34 | Good |
| Delay/ms | 275 | 217 | 221 | Bad |
| Jitter/ms | 183 | 116 | 126 | Bad |

For delay and jitter, both the basic CCN and *PIT probe* are greater than 125 milliseconds. According to the literature [15], they belong to the "Bad" QoS category. *PIT probe* performs similarly to the other two methods in terms of delay. However, it has a jitter of 116 milliseconds, which is less than 125 milliseconds. Thus, it belongs to the "Good" QoS category. This shows that the network delay variation of *PIT probe* is more stable than the basic CCN and *FIB probe*. *PIT probe* is superior to the basic CCN and *FIB probe* in terms of delay and jitter. This is consistent with the results shown in Figures 6 and 7. *PIT probe* directly reduces customers' waiting time by selecting the most popular PIT entry as a probe to find the best content provider. Hence, *PIT probe* gets superior performance on response time than other approaches.

Although basic CCN and our approach are in the same QoS category. However, our approach performs better in terms of real values.

4.4.6. The Impact of Router Failures and Traffic Variations

Figure 8 illustrates the impact of router failures on the performance of our approach, RECIF+PIF [41] and EEGPR [31]. This simulation adopts PoP-level SPRINT topology [7,33]. To simulate the failures of routers, several routers are randomly selected to crash after running half of the simulations. Furthermore, the centrality threshold and retransmission threshold of RECIF+PIF has been set to 9 and 5, respectively. All results are the average of five simulations. As demonstrated in Figure 8a,b, as the number of router failures increases, so does network delay and packet loss. In particular, the delays of RECIF+PIF and EEGPR are 5.4 milliseconds and 9.9 milliseconds longer than that of the *PIT probe*, respectively. EEGPR only improves the routing table for data packets while RECIF+PIF focuses on decreasing the number of interests instead of improving the routing table to speed up routing.

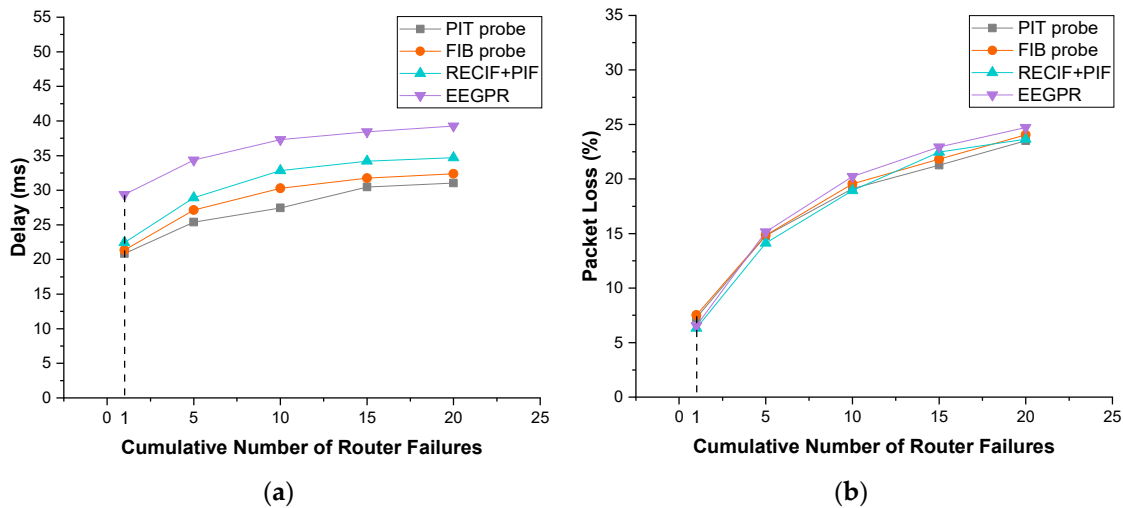

**Figure 8.** Impact of router failures. (**a**) Delay; (**b**) packet loss.

Figure 9 illustrates the impact of traffic variations of our approach, RECIF+PIF and EEGPR. The simulating parameters follow the settings of the router failure experiment. The traffic variation was simulated by increasing the frequency of interest packets sent by consumers. The frequency was set from 1 to 30 interests per second. The bandwidth was set to 1024 bits/s. As shown in Figure 9, traffic variations are inversely proportional to delay and packet loss for all the approaches. In particular, comparing the delays of 1 and 30 router failures, the increments of other approaches are 1.07 milliseconds and 3.95 milliseconds longer than the *PIT probe*, respectively. That is because, with the increase of interest packet frequency, more probes are used to update FIB so the accuracy and expiration of FIB are improved.

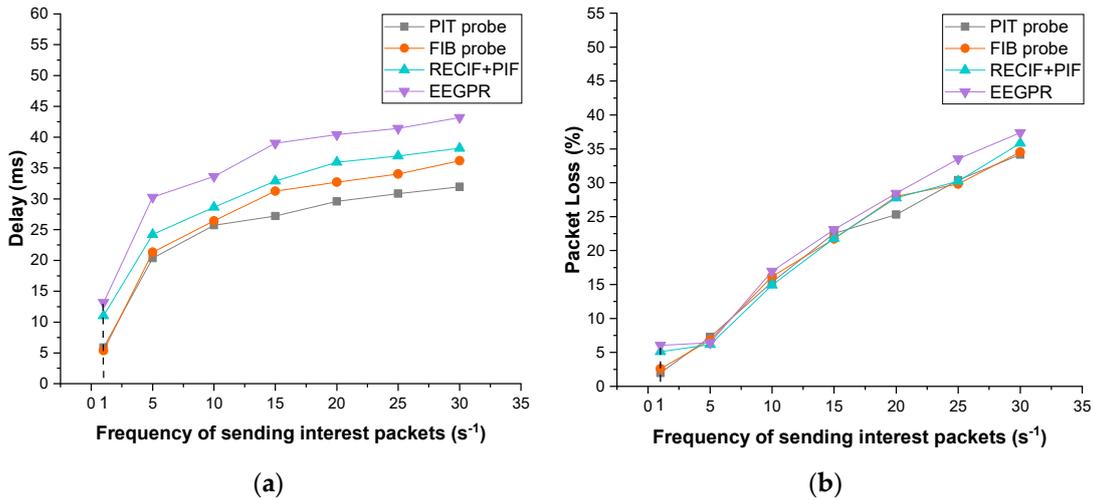

**Figure 9.** Impact of traffic variations. (**a**) Delay; (**b**) packet loss.

To summarize, the results demonstrate that our probe-based approach performs better than RECIF+PIF and EEGPR. Owing to speedup FIB updating by probes, our approach provides more reliable interest packet routing when the router failures.

## 5. Conclusions

This paper presents an efficient *probe*-based routing algorithm for content-centric networking that real-time explores in-network caching to ensure the routing table storing the optimal paths to the nearest content provider up to date. The *probe*-based approach modifies data structures of basic CCN, including packet formats and routing tables. Compared to other content exploration methods that relied on extra designated packets that increase the network burden, our approach minimizes the cost during the exploration process. The *probe*-based routing table updating algorithm designed in this paper improves the efficiency of FIB. Several *probe* selections are simulated to demonstrate how to choose the probe and compared with basic CCN on different performance measurements. In terms of qualitative analysis, compared to the basic CCN, the additional computational overhead of our approach is $O(N_{CS} + N_{rt} + N_{FIB} * N_{SPT})$ and $O(N_{FIB})$ on processing interest packets and data packets, respectively. However, the quantitative experimental results show that our approach reduces the number of timeout interest by 6% and the average response time by 0.6 s. In addition, the QoS category of basic CCN and our approach are analyzed based on measurement results of throughput, packet loss, delay, and jitter. Although basic CCN and our approach belong to the same QoS category, our approach outperforms others in terms of real values. Furthermore, the impact of router failures and traffic variations on the performance of our approach, RECIF+PIF and EEGPR are simulated and compared on more complicated network topology. The results demonstrate that our probe-based approach performs better than RECIF + PIF and EEGPR. Owing to speedup FIB updating by probes, our approach provides more reliable interest packet routing when the router failures.

In summary, the results reveal that the *probe*-based approach raises the accuracy of the routing table, and reduces the network congestion of basic CCN and response time of interests to ensure more efficient routing.

**Funding:** This work was supported in part by the grant from The Ministry of Science and Technology of Taiwan and the Industrial Technology Research Institute of Taiwan, under grant Nos. MOST 108-2221-E-006 -095 -MY2 and MOST 110-2221-E-006 -008 -.

**Institutional Review Board Statement:** Not applicable.

**Informed Consent Statement:** Not applicable.

**Data Availability Statement:** Not applicable.

**Conflicts of Interest:** The authors declare no conflict of interest.